\documentclass[a4paper]{article}

\usepackage{INTERSPEECH2020}
\usepackage{subfigure}

\title{Weak-Attention Suppression For Transformer Based Speech Recognition}
\name{Yangyang Shi, Yongqiang Wang, Chunyang Wu, Christian Fuegen, Frank Zhang, Duc Le, Ching-Feng Yeh, Michael L. Seltzer}
\address{Facebook AI, USA}
\email{\{yyshi,yqw,chunyang,fuegen,frankz,duchoangle,cfyeh,mikeseltzer\}@fb.com}

\begin{document}

\maketitle
\begin{abstract}

Transformers, originally proposed for natural language processing (NLP) tasks, have recently achieved great success in automatic speech recognition (ASR). However, adjacent acoustic units (i.e., frames) are highly correlated, and long-distance dependencies between them are weak, unlike text units. It suggests that ASR will likely benefit from sparse and localized attention. In this paper, we propose Weak-Attention Suppression (WAS), a method that dynamically induces sparsity in attention probabilities. We demonstrate that WAS leads to consistent Word Error Rate (WER) improvement over strong transformer baselines. On the widely used LibriSpeech benchmark, our proposed method reduced WER by $10\%$ on \texttt{test-clean} and $5\%$ on \texttt{test-other} for streamable transformers, resulting in a new state-of-the-art among streaming models. Further analysis shows that WAS learns to suppress attention of non-critical and redundant continuous acoustic frames, and is more likely to suppress past frames rather than future ones. It indicates the importance of lookahead in attention-based ASR models.
\end{abstract}


\section{Introduction}
In recent years, models based on transformers \cite{vaswani_2017} and their variants have achieved state-of-the-art results in many natural language processing (NLP) tasks, such as language modeling, machine translation, and natural language understanding \cite{dai_2019,devlin_2018,Raffel_2019}. The transformer relies on the multi-head self-attention method, eschewing the recurrent connection in recurrent neural networks. The attention method connects arbitrary positions in the whole sequence directly, allowing the model to capture long-range dependencies regardless of distance. Each transformer layer performs computations for the whole sequence in parallel, thus improving efficiency.

In automatic speech recognition (ASR), transformer-based architectures have also showed superior performance across various modeling paradigms, including sequence-to-sequence \cite{dong2018speech,Karita_2019,sperber2018self,zhou2018syllable}, neural transducer \cite{zhang_2020,Yeh_2019}, as well as traditional hybrid \cite{povey2018time,yongqiang_2019_icassp} and CTC \cite{Salazar2019_ICASSP} systems.

While similar in some aspects, ASR tasks are very different from many NLP tasks. ASR systems extract the input features from continuous signals rather than discrete text units. Many acoustic units (i.e., frames) are typically needed to convey a single text unit's semantic meaning. Many of these acoustic units are not critical in long-range dependencies, for example, silence. Based on these observations, we propose the Weak-Attention Suppression (WAS) method to improve transformer-based models for ASR. The method aims to induce sparsity in the attention probability distribution by dynamically determining a threshold for each time frame. All attention probabilities smaller than this threshold are set to zero, and the remaining probabilities are re-normalized to sum to one.

To verify the performance of the proposed method, we carry out experiments on the widely used LibriSpeech corpus \cite{Panayotov2015}. WAS consistently improves Word Error Rate (WER) of transformer-based hybrid acoustic models across different model sizes (12-layer 40M parameters vs. 24-layer 81M parameters) and variants (non-streamable full context \cite{yongqiang_2019_icassp} vs. streamable limited context \cite{chunyang_2020_interspeech}). The relative WER improvement ranges from $6\%$ to $10\%$ on \texttt{test-clean} and $2\%$ to $6\%$ on \texttt{test-other}, with larger impact on streamable transformers. Our results with WAS and streamable transformers established a new state-of-the-art on Librispeech among streaming models to the best of our knowledge.


A more in-depth analysis of individual utterances and the whole dev-clean split in LibriSpeech data shows that WAS suppresses attention from non-critical acoustic units, such as silence. For adjacent acoustic units, attention from redundant units is suppressed, especially for the lower layers. The analysis also demonstrates that past acoustic units are more likely to be suppressed than future ones. In other words, lookahead is especially crucial for transformer-based acoustic models.

\section{Related Work}
The proposed WAS method aims to improve transformer-based acoustic models in hybrid speech recognition for both streaming \cite{povey2018time,chunyang_2020_interspeech}, and non-streaming \cite{yongqiang_2019_icassp} applications. 

For the non-streaming case, we use the same model architecture proposed in \cite{yongqiang_2019_icassp}, where the whole model architecture consists of three parts, convolution layers, transformer layers, and auxiliary intermediate layer losses. The convolution layers play a similar role as another positional encoding method \cite{mohamed2019transformers}. A stack of transformers with both multi-head self-attention and feed-forward network (FFN) is used on top of the convolution layers. To effectively train the deep transformers, the final loss function is interpolated with auxiliary intermediate layer losses \cite{Andros2019} that pass the gradients to the intermediate layers in the deep structure. 

To extend transformer-based acoustic models for streaming applications, the work \cite{chunyang_2020_interspeech} applies block processing to segment the whole utterance into multiple chunks with lookahead context. To carry over information across chunks, they modify the self-attention with the augmented memory bank. Each slot in the augmented memory bank stores the embedding vectors for previous chunks.

Sparse attention in the transformer has been explored in \cite{Kitaev_2020} to reduce model inference complexity. Unlike this work, our method aims to improve ASR accuracy by inducing sparsity in both training and inference. 

Different from the works in \cite{Martins_2016,Correia_2019} that achieve sparse attention by replacing the softmax function with $\alpha$-entmax or sparsemax, our method keeps the softmax function and suppresses the attention based on the probability distribution. Comparing these different methods for suppressing attention will be an interesting topic for future research. 

\section{Transformer Based Acoustic Model With Weak-Attention Suppression}
Before explaining the details of the WAS, the following subsection gives a short description of multi-head self-attention.

\subsection{Multi-Head Self Attention}
Given the input embedding sequence $\boldsymbol{X} \in \mathbb{R}^{L \times d_i}$ with sequence length $L$, the projection matrices $\mathbf{W}_{\rm q}$, $\mathbf{W}_{\rm k}$ and $\mathbf{W}_{\rm v}$ transform the $\boldsymbol{X}$ into \emph{query}, \emph{key} and \emph{value} space, respectively.
\begin{align}
   \boldsymbol{Q}=\boldsymbol{X}\mathbf{W}_{\rm q}, \boldsymbol{K}=\boldsymbol{X}\mathbf{W}_{\rm k},
   \boldsymbol{V}=\boldsymbol{X}\mathbf{W}_{\rm v}.
\end{align}
The attention probabilities are computed in dot-product way as follows:
\begin{align}
    \boldsymbol{A} = {\rm softmax}(\frac{\boldsymbol{K}\boldsymbol{Q}^T}{\sqrt{d_i}}),
\end{align}
where $\boldsymbol{A}$ is a $L \times L$ matrix. Each element $\alpha_{i,j}$ represents the attention probability of the \emph{query} at $i$ position with the \emph{key} at $j$ position.

Given $\boldsymbol{A}$, the output embedding sequence of self-attention is obtained via:
\begin{align}
    \boldsymbol{Z} = \mathrm{Dropout}(\boldsymbol{A})\boldsymbol{V}.
\end{align}
Rather than performing a single self-attention function with queries, keys and values, it was found beneficial to run self-attention $h$ times in parallel with different projection matrices $\mathbf{W}_{\rm q}$, $\mathbf{W}_{\rm k}$ and $\mathbf{W}_{\rm v}$. The outputs from each self-attention are concatenated and linearly projected, resulting to final output, i.e., 
\begin{align}
    \boldsymbol{O}  = \mathbf{W}_{\rm o}\mathrm{Concate}(\boldsymbol{Z}^{1},...,\boldsymbol{Z}^{h})
\end{align}
where $\mathbf{W}_{\rm o} \in \mathbb{R}^{d_i \times hd_{\rm v}}$, and $\boldsymbol{Z}^{i}$ is the output from the $i$-th self-attention. 

\subsection{Weak-Attention Suppression}
In speech recognition tasks, many acoustic units, e.g., silence, may not play a critical role in long-distance dependencies. For continuous acoustic units that share similarities, some of them may be redundant for long-distance dependencies. Comparing with many NLP tasks, the number of acoustic units in audio utterances are much more significant than the number of text units in sentences. Due to these factors, the sparse attention is more desirable for speech recognition. However, in self-attention, the softmax function is used to get the attention probabilities. One limitation of softmax function is that it always generates dense attention, i.e., $\mathrm{softmax}(X)_i\neq0$; all the elements of the softmax function is over zero.   

The WAS sets the attention probabilities smaller than a threshold to zero and normalizes the rest attention probabilities. The threshold is determined based on the following schema.
\begin{equation} \label{eq1}
\theta_i = m_i - \gamma\delta_i 
\end{equation}
where $m_i$ and $\delta_i$ are the mean and standard deviation of the attention probability for $i$-th position in \emph{query}, respectively. So the threshold $\theta_i$ can be represented as 
\begin{align}
    \theta_i  = \frac{1}{L} - \gamma\sqrt{\frac{\sum_{j=1}^{L}(\alpha_{i,j}-\frac{1}{L})^2}{L-1}},
    \label{theta}
\end{align}
where $L$ is the length of \emph{key} in self-attention. The attention probability $\alpha_{i,j}$ less than $\theta_i$, is set to zero. The rest non-zero attention probabilities are re-normalized. 

In practical implementation, we realize the re-normalization in two steps. Firstly, using softmax function on attention weights, we get the attention probabilities. Based on formula (\ref{theta}), we replace the attention weight to negative infinity when its attention probability is lower than the threshold. Applying the softmax function on the processed attention weight generates the re-normalized attention probabilities. 

\section{Experiments}

\subsection{Data}
To verify the performance of the proposed method, we conduct experiments on the LibriSpeech corpus \cite{Panayotov2015}. LibriSpeech is an open-source speech corpus that contains 1000 hours of speech derived from audiobooks in the LibriVox project. The audio data \footnote{http://www.openslr.org/12/}, language model data, and pre-trained language models \footnote{http://www.openslr.org/11/} are available for downloading. The development data and evaluation data in LibriSpeech are split into simple ``clean" subsets and more difficult ``other" subsets. We use the official 4-gram language model for decoding in all experiments.


\subsection{Setup}
In all experiments, we use context and positional dependent graphemes (i.e., chenones) as output units \cite{le2019senones}. We bootstrap the HMM-GMM system following the standard Kaldi \cite{Povey_ASRU2011} LibriSpeech recipe. For feature extraction, we use 80-dimensional logMel filter bank features at a 10ms frame rate. To increase the training robustness, both speed perturbation \cite{ko2015audio} and \emph{SpecAugment} \cite{park2019specaugment} without time warping are used. 

We apply the same model architecture as \cite{yongqiang_2019_icassp,chunyang_2020_interspeech}, which has two VGG blocks \cite{simonyan2014very} followed by a stack of transformer layers. Each VGG block has two consecutive 3-by-3 convolution layers followed by a Relu activation function and a max-pooling layer. While the first VGG block uses 32 channels, the second VGG block uses 64 channels. Both VGG blocks use a 2-by-2 Max-pooling. The first VGG block uses stride 2; the second one uses stride 1. Overall, two VGG blocks have stride 2. From an input sequence of 80-dim feature vector at a 10ms rate, the VGG blocks generate a 2560-dim feature vector sequence at a 20ms rate. 

Each transformer layer applies 8 heads of self-attention. The dimensionality of input and output for each transformer layer is 512, and the inner-layer of FFN has dimensionality 2048. We experiment with different model sizes (12 transformer layers, and 24 transformer layers architectures) to verify the improvement from the WAS. To overcome the training divergence issue for deep transformer models, e.g., 24 transformer layers, we apply an auxiliary incremental loss \cite{Andros2019}. A linear transformation, together with a Relu function, projects the outputs from the 6/12/18-th transformer layers to the final output space. The auxiliary incremental loss takes the projected outputs to compute the auxiliary CE losses interpolated with the original CE loss with a 0.3 weight.  

In all experiments, we use the adam optimizer \cite{kingma2014adam} with a tri-stages learning-rate strategy, i.e., warming-up stage, holding stage, and decaying stage. 8K updates in the warming-up stage increase the learning rate from 1e-5 to 1e-3 for non-streaming transformer-based models and 3e-4 for streaming transformer-based models, respectively. The holding stage uses 100K updates. The decaying stage exponentially decreases the learning rate to 1e-5.
 
To efficiently use GPU resources, the batch size is dynamically determined. Each batch contains around 10,000 to 20,000 frames, including padding frames. We train all the models using 32 Nvidia V100 GPUs. To train the transformer-based model effectively, we segment the training utterances into less than 10 seconds using forced alignment results from an existing latency-controlled BLSTM acoustic model. We select the best model based on WER on the dev set and report its result on test data.

\subsection{Results}

\begin{table}[htb]
    \centering
    \begin{tabular}{|c|c|cc|}
    \hline
    Model Arch     & \#Params (M) & \rm{test-clean} & \rm{test-other} \\
    \hline\hline
    BLSTM\cite{yongqiang_2019_icassp}   & 79 & 3.11 & 7.44 \\
    vggBLSTM\cite{yongqiang_2019_icassp} & 95 & 2.99 & 6.95 \\
    \hline
    vggTrf-12 & 40 & 3.11 & 7.14 \\
    vggTrf-12-WAS & 40 & 2.93 & 6.73 \\
    \hline
    vggTrf-24\cite{yongqiang_2019_icassp} & 81 & 2.66 & 5.64 \\
    vggTrf-24-WAS & 81 & 2.50 & 5.55 \\
    \hline
    AmTrf-24\cite{chunyang_2020_interspeech} & 81 & 3.09 & 7.08 \\
    AmTrf-24-WAS & 81 & 2.78 & 6.71 \\
    \hline
    \end{tabular}
    \caption{WER comparison on the LibriSpeech benchmark. For models with citations, we directly use the results from the referred paper. `-WAS' means the weak-attention suppression with $\gamma=0.5$ is applied. Note results in this table are obtained without the NNLM rescoring.}
    \label{librispeech}
\end{table}

Table~\ref{librispeech} gives the WER comparison of different hybrid models on LibriSpeech data. Overall, the transformer-based models show improvement over recurrent neural network counterparts. For small transformer based model `vggTrf-12`, applying WAS brings a relative WER reduction of $5.8\%$ on test-clean and $5.7\%$ on test-other. On a large model with 24 transformer layers, WAS generates a relative WER reduction of $6.0\%$ on test-clean and $5.7\%$ on test-other. `AmTrf-24' is a streaming transformer-based model using augmented memory that limited the attention within an audio segment with 320ms lookahead. The WAS achieves relative WER reduction $10.0\%$ on test-clean and $5.2\%$ on test-other over the streaming baseline.

\begin{table}[htb]
    \centering
    \begin{tabular}{|c|c|cc|}
    \hline
    Model & $\gamma$ & test-clean & test-other   \\
    \hline
    vggTrf-12 &- & 3.11 & 7.14 \\
    vggTrf-12-WAS & 0.0  & 2.97 & 7.13 \\
    vggTrf-12-WAS & 0.5  & 2.93 & 6.73 \\
    vggTrf-12-WAS & 1.0  & 3.04 & 6.92 \\
    \hline
    \end{tabular}
    \caption{Word error rate comparison of different $\gamma$ selection in weak-attention suppression on LibriSpeech data.}
    \label{tab:gamma}
\end{table}

Table~\ref{tab:gamma} gives the WER results from different $\gamma$ selection on LibriSpeech data using the small size of the transformer model. $\gamma=0.5$ gives us the best results. We constrain the $\gamma$ selection from 0 to 1. If the attention probability is close to a normal distribution, we suppress roughly $16\%$ to $50\%$ the attention. According to our analysis, when $\gamma=0.5$, $36\%$ of attention got suppressed in the first transformer layer.

\subsection{Attention Suppression Analysis}
In this section, we analyze the effect of WAS in one individual utterance and the whole dev-clean split in LibriSpeech using the model `vggTrf-12-WAS` in Table \ref{librispeech}.

In individual utterance analysis, we average the portion of the suppressed attention over the multiple heads and the whole input sequence for each transformer layer. The following formula represents the processing.
\begin{align}
s_{i,j}^{k} = \left\{ \begin{array}{cc} 
                1 & \hspace{5mm} \alpha_{i,j}^k < \theta_{i}^{k} \\
                0 & \hspace{5mm} \alpha_{i,j}^k \geq \theta_{i}^{k} \\
                \end{array} \right.
\end{align}
where $\alpha_{i,j}^k$ represents the attention probability of the $i$-th position in \emph{query} with $j$-th position in \emph{key} from the $k$-th head. $\theta_{i}^k$ is the suppression threshold for $i$-th position in \emph{query} in the $k$-th head. The threshold is determined according to formula~(\ref{theta}).
\begin{align}
f(j) = \frac{\sum_{i=1}^{i=L}\sum_{k=1}^{k=H}s_{i,j}^{k}}{LH}
\end{align}
where $L$ is the length of \emph{key} and $H$ the number heads in multi-head self-attention. The function value $f(j)$ indicates the weakness of the attention at the $j$-th position in the \emph{key}. 

\begin{figure}[t!]
    \subfigure[wav]
    {
        \includegraphics[width=3.0in]{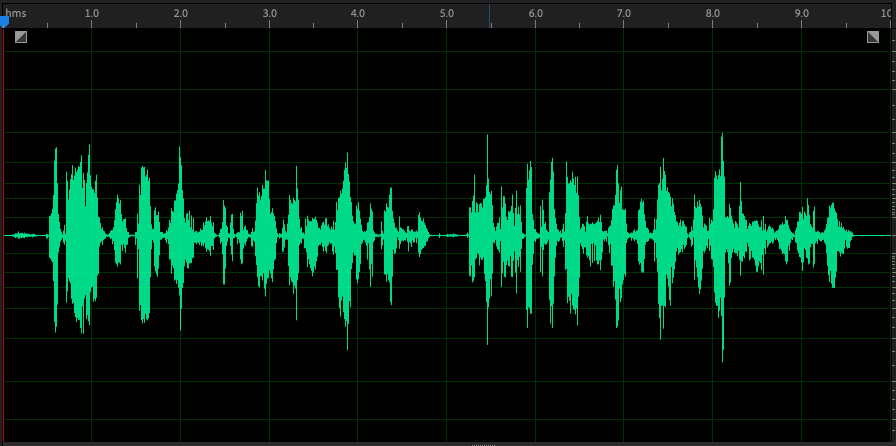}
        \label{fig:wav}
    }
    \\
    \subfigure[layer 1]
    {
        \includegraphics[width=3.0in]{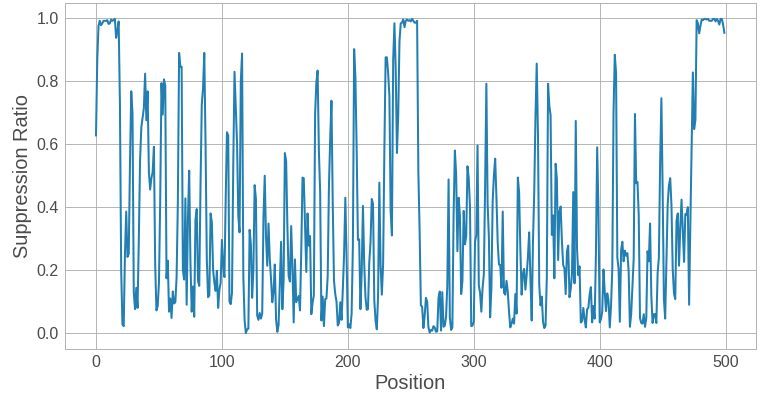}
        \label{fig:layer_1}
    }
    \\
    \subfigure[layer 6]
    {
        \includegraphics[width=3.0in]{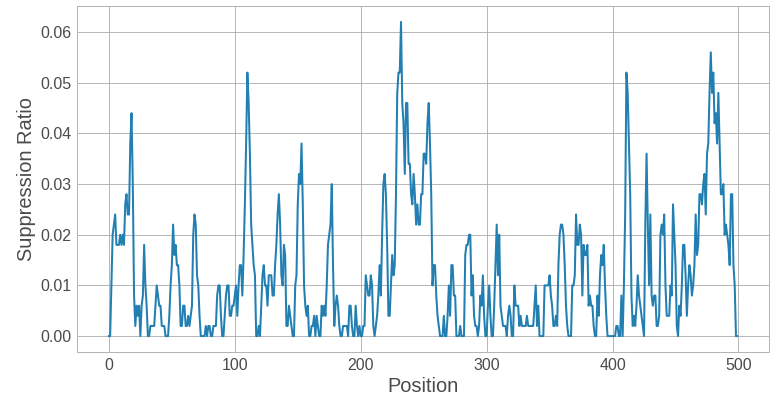}
        \label{fig:layer_6}
    }
    \\
    \subfigure[layer 12]
    {
        \includegraphics[width=3.0in]{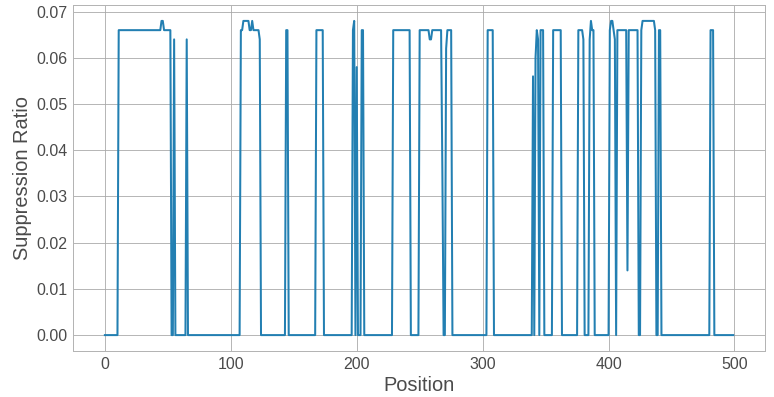}
        \label{fig:layer_12}
    }
    \caption{Attention suppression analysis for one utterance in dev-clean using `vggTrf-12-WAS` in Table \ref{librispeech}. (a) gives the audio wave of the utterance. (b), (c), (d) represent the average attention suppression portion for different positions in the input sequence at the 1-st, 6-th, 12-th transformer layer, respectively.}
    \label{fig:one_utterance}
\end{figure}

Figure~\ref{fig:layer_1},\ref{fig:layer_6} and \ref{fig:layer_12} draws the function $y=s(j)$ for 1st, 6-th and 12-th transformer layer. The horizontal axis stands for the position in the \emph{key}. The vertical axis stands for the averaged portion of suppressed attention. Note `vggTrf-12-WAS` uses VGG blocks with stride 2. For one-second audio using frameshift 10ms, the length of the input sequence to each transformer layer is 500.

By comparing the audio wave \ref{fig:wav} with \ref{fig:layer_1}, it is obvious to see there are three peaks (at the beginning, in the middle and at the end) in \ref{fig:layer_1} that are the exact position of a long silence in the audio wave. The rest of the small peaks in \ref{fig:layer_1} seem to be in the rough location of short silence in the audio wave in \ref{fig:wav}. This phenomenon indicates that using weak-attention suppression in the first transformer layer can suppress the attention to silence.

Zooming into Figure \ref{fig:layer_1} and Figure \ref{fig:layer_6}, there are dramatic ups-and-downs even for the non-silence consecutive positions in the input sequence that are supposed to share acoustic similarities. This phenomenon suggests that for consecutive frames that share acoustic similarities, the weak attention suppress method suppresses the redundant information for attention.

Comparing the Figures \ref{fig:layer_1}, \ref{fig:layer_6} and \ref{fig:layer_12} vertically, there are less fluctuations with higher layer. This observation may suggest that the lower transformer's self-attention can capture the small difference for consecutive frames while the higher layer captures the salient invariant features for consecutive frames.  

\begin{figure}[t!]
    \subfigure[position 200 layer 1]
    {
        \includegraphics[width=1.5in]{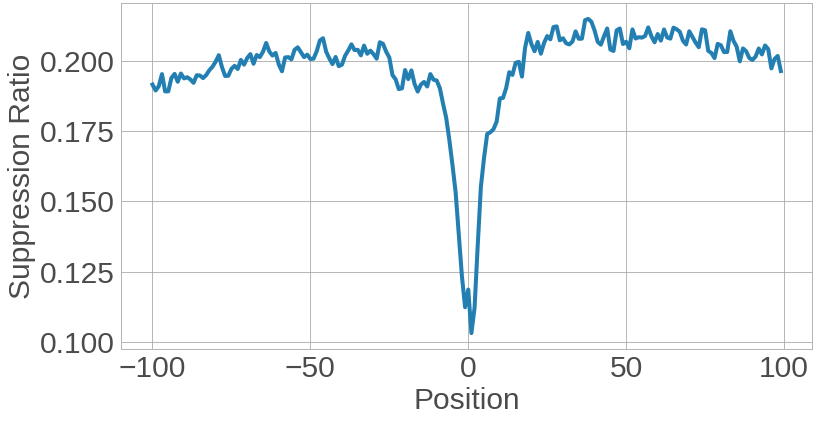}
        \label{fig:200_layer1}
    }
    \subfigure[position 200 layer 12]
    {
        \includegraphics[width=1.5in]{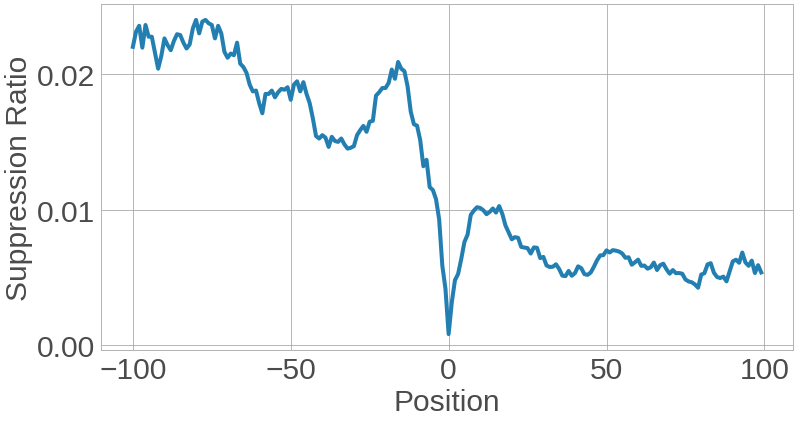}
        \label{fig:200_layer12}
    }
    \\
    \subfigure[position 400 layer 1]
    {
        \includegraphics[width=1.5in]{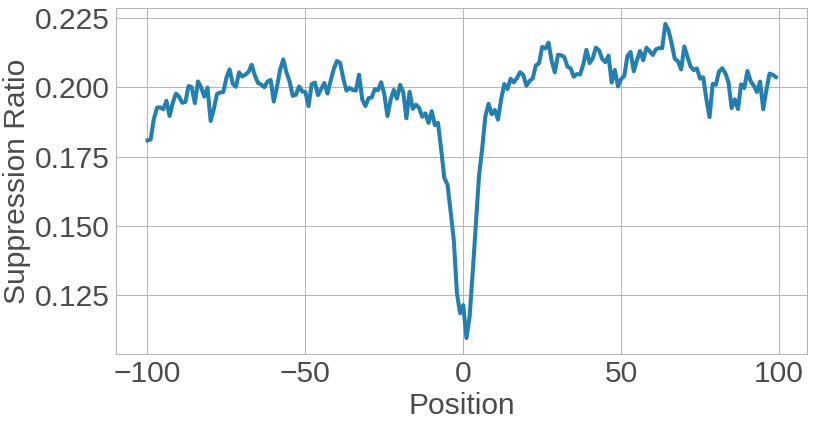}
        \label{fig:400_layer1}
    }
    \subfigure[position 400 layer 12]
    {
        \includegraphics[width=1.5in]{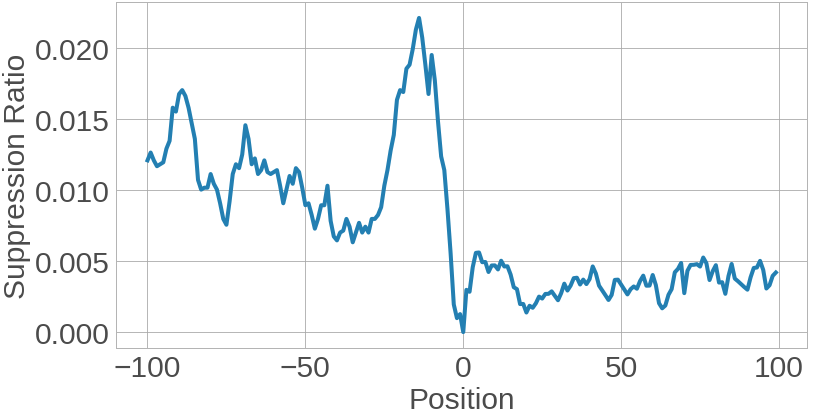}
        \label{fig:400_layer12}
    }
    \\
    \subfigure[position 600 layer 1]
    {
        \includegraphics[width=1.5in]{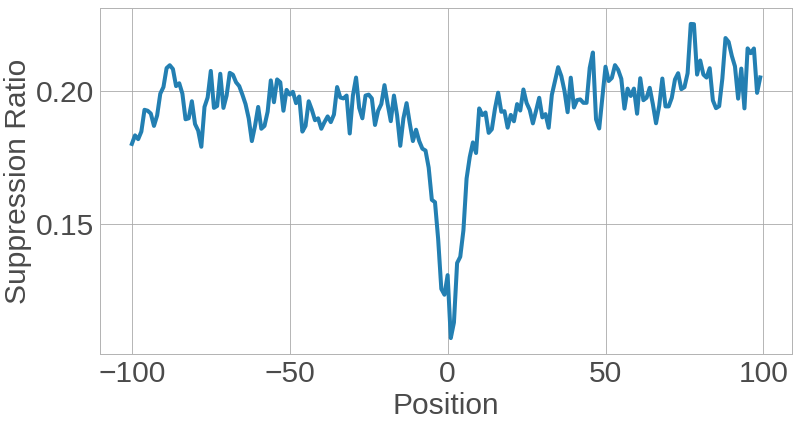}
        \label{fig:600_layer1}
    }
    \subfigure[position 600 layer 12]
    {
        \includegraphics[width=1.5in]{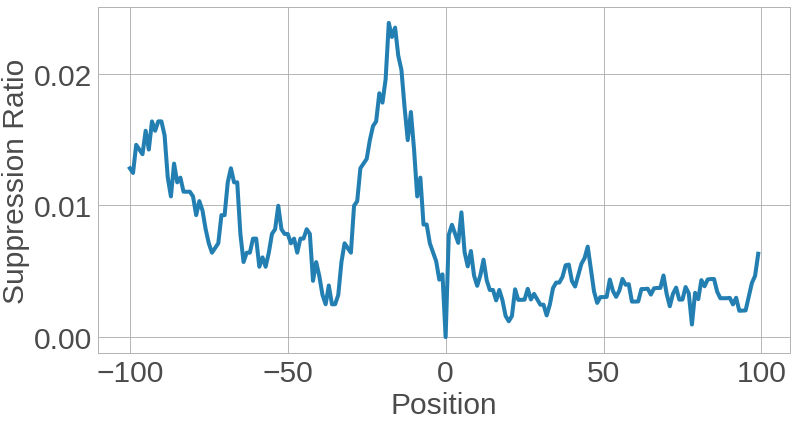}
        \label{fig:600_layer12}
    }
    \caption{Attention suppression analysis over the dev-clean using `vggTrf-12-WAS` in Table \ref{librispeech}. Each sub-figure gives the average portion of attention suppressed at a specific position in a specific layer within the window size 200. We take the average over different heads in multi-head self-attention and different utterances over the dev-clean dataset.}
    \label{fig:whole_dev_clean}
\end{figure}

Figure~\ref{fig:whole_dev_clean} gives the weak-attention suppression analysis at different positions from different transformer layers. For a specific position $i$, we average the portion of the suppressed attention over multiple heads and different utterances in the dev-clean dataset. We represent the process as follows:
\begin{align}
f_i(j) = \frac{\sum_{n=1}^{n=N}\sum_{k=1}^{k=H}s_{i,j}^{k,n}}{NH}
\end{align}
where $s_{i,j}^{k,n}$ means whether to suppress the attention between $i$-th position in \emph{query} with $j$-th position in \emph{key} from $k$-th head in utterance $n$. $N$ is the number of utterances in dev-clean data and $H$ the number heads in multi-head self-attention. The function value $f_i(j)$ indicates the weakness of the attention at the $j$-th position in the \emph{key} for $i$-th position in \emph{query}. For easy illustrating, we only draw the curve within window size 200 with 100 left and 100 right contexts.

Figure~\ref{fig:whole_dev_clean} reveals the following three phenomenons. Firstly, the left column in the figure shows that more attention gets suppressed in the lower layer when the attention distance gets further. The deep valley of attention suppression happens in the context window of -5 to 5, i.e., 100 ms left context and 100 ms right context. The right column shows that in the 12-th layer, more attention gets suppressed from the left context than from the right context. One common phenomenon in both columns is that there are roughly 10 times more attention gets suppressed in the 1-st layer than the 12th layer.

\section{Conclusions}
In this paper, we proposed a weak-attention suppression (WAS) method to address the sparse attention for speech recognition. The attention probability smaller than a dynamic estimated threshold was zeroed out by WAS. The analysis showed several exciting phenomena of the proposed method. In the lower transformer layer, the attention suppression happens for the silence and redundant acoustic units. Even using the same attention suppression schema, less attention got suppressed in the upper layer than the lower layer. The further the dependency distance was, the more attention got suppressed. In the top transformer layer, the WAS suppressed more attention from the left context than the right context. Experiments on LibriSpeech showed that the WAS improves the transformer-based model with different model sizes in both streaming and non-streaming scenarios. Especially for a streamable transformer-based acoustic model, the proposed WAS got relative word error rate reduction by $10.0\%$ on \texttt{test-clean} and $5.2\%$ on \texttt{test-other}. 
\bibliographystyle{IEEEtran}
\bibliography{myref}


\end{document}